\documentclass{kapproc} % Computer Modern font calls
\usepackage{procps} 
\usepackage[dvips]{graphicx}
\upperandlowercase
\kluwerbib % will produce this kind of bibliography entry:
\def\gtsim{\lower.5ex\hbox{$\buSildrel > \over\sim$}}
\def\ltsim{\lower.5ex\hbox{$\buildrel < \over\sim$}}
\def\arcsec{^{\prime\prime}}
\def\arcmin{^\prime}

\def\etal{{\it et al.~}}
\def\deg{^\circ}

\def\vb{$V$-band}

\def\zb{$z$-band}
\def\apjl{ApJL}
\def\apj{ApJ}
\def\apjs{ApJS}
\def\mnras{MNRAS}

\def\aap{A\&A}
\def\aaps{A\&A Suppl.}

\begin{document}

\articletitle
[Evolution and Impact  of Bars over the last 8 Gyr: 
Early Results from GEMS]
{Evolution and Impact  of Bars over the \\
Last Eight Billion Years: Early Results from GEMS
}

\author{
Shardha Jogee\altaffilmark{1}, 
Fabio D. Barazza \altaffilmark{1},
Hans-Walter Rix \altaffilmark{2},
James Davies \altaffilmark{1},
Inge  Heyer \altaffilmark{1},
Marco Barden\altaffilmark{2},
Steven V.W. Beckwith \altaffilmark{1},
Eric F. Bell\altaffilmark{2},
Andrea Borch\altaffilmark{2},
John A. R. Caldwell\altaffilmark{1},
Christopher Conselice\altaffilmark{3},
Boris H\"{a}ussler\altaffilmark{2},
Catherine  Heymans\altaffilmark{2},
Knud Jahnke\altaffilmark{4},
Johan H. Knapen\altaffilmark{5},
Seppo Laine\altaffilmark{6},
Gabriel M. Lubell\altaffilmark{7},
Bahram Mobasher \altaffilmark{1},
Daniel H. McIntosh\altaffilmark{8}
Klaus Meisenheimer\altaffilmark{2},
Chien Y. Peng\altaffilmark{9},
Swara Ravindranath\altaffilmark{1},
Sebastian F. Sanchez\altaffilmark{4},
Isaac Shlosman\altaffilmark{10},
Rachel S. Somerville\altaffilmark{1},
Lutz Wisotzki\altaffilmark{4} and 
Christian Wolf\altaffilmark{11}
} 
\affil{
\altaffilmark{1} Space Telescope Science Institute, 
3700 San Martin Drive, Baltimore MD 21218, U.S.A, 
\altaffilmark{2} 
Max-Planck Institute for Astronomy, D-69117 Heidelberg, Germany,
\altaffilmark{3} 
Caltech, Dept. of Astronomy, Pasadena, CA 91125, U.S.A,
\altaffilmark{4} 
Astrophysikalisches Institut Potsdam, D-14482 Potsdam, Germany,
\altaffilmark{5} 
University of Hertfordshire, Hatfield, Herts AL10 9AB, UK
\altaffilmark{6} 
Spitzer Science Center, Caltech, Pasadena, CA 91125, U.S.A,
\altaffilmark{7} 
Vasser College, Dept. of Physics and Astronomy, Poughkeepsie, NY 12604, U.S.A,
\altaffilmark{8} 
University of Massachusetts, Amherst, MA 01003, U.S.A,
\altaffilmark{9} 
University of Arizona, Tucson, AZ 85721, U.S.A,
\altaffilmark{10} 
University of Kentucky, Lexington, KY 40506-0055, U.S.A,
\altaffilmark{11}
University of Oxford Astrophysics, Oxford OX1 3RH, UK
}

\begin{abstract}
Bars drive the dynamical evolution of disk galaxies 
by redistributing mass and angular momentum, and they 
are ubiquitous in present-day spirals.
Early studies  of the Hubble Deep Field  reported 
a dramatic decline in the 
rest-frame optical bar fraction $f_{\rm opt}$ 
to below 5\%   at  redshifts $z >$~ 0.7, implying that disks 
at these epochs are fundamentally different from present-day spirals.
The GEMS bar project, based on $\sim$ 8,300 galaxies with $HST$-based
morphologies and accurate redshifts  over the range 0.2--1.1,  
aims at constraining the evolution and  impact of bars 
over the last 8 Gyr. We present early results  indicating 
that   $f_{\rm opt}$  remains $\sim$ constant at  
$\sim$ 30\%  over $z \sim$~0.2--1.1, 
corresponding to lookback times of  $\sim$~2.5--8 Gyr.
The bars detected at $z>$~0.6  are primarily strong with 
ellipticities of  0.4--0.8. Remarkably, 
the  bar fraction  and  range of bar sizes observed  
at $z>$~0.6,  appear to be  comparable to the values measured 
in the local Universe for bars of corresponding strengths. 
Implications for bar evolution models are discussed.
\end{abstract}


%\footnote {Here is a sample footnote 
% which will normally format as an endnote at the end of the article.}

%\section[Introduction] 
%{Introduction}

\section[Open issues in bar-driven galaxy evolution] 
{Open issues in bar-driven galaxy evolution} 

Large-scale bars are ubiquitous in present-day spirals, with reported 
optical bar fractions of $\sim$ 30 \% for both strong and 
weak bars (e.g., Eskridge \etal 2002aa, hereafter ES02; $\S$ 3). 
Such bars can  drive the dynamical evolution of disk galaxies 
by redistributing mass and angular momentum. In particular, they 
are believed  to efficiently  drive gas from the outer disk of galaxies  
into the inner kpc via  gravitational torques. 
% Gravitational torques  operate on  a timescale  
% comparable to the orbital timescale and  provide, therefore,  
%the most efficient  way  of reducing  angular momentum 
% on large to intermediate scales (tens of kpc -- a few 100 pc.
%A  large-scale bar efficiently drives gas from the outer disk 
%The bar-driven gas flow slows or 
% even stalls as it crosses the inner Lindblad  resonance (ILR) 
There is mounting observational evidence for bar-driven gas inflow.
%Moderate ($4\arcsec$) resolution  
Observations of cold or ionized gas velocity fields show   evidence 
for shocks and non-circular  motions along  the large-scale 
stellar bar  (e.g., Regan, Vogel, \& Teuben 1997). 
Barred galaxies  show shallower metallicity gradients across 
their galactic disks  than  unbarred ones (Martin \& Roy 1994). 
CO($J$=1--0) interferometric surveys show that, on average,
the  molecular gas central  concentration  in the inner kpc 
is higher  in barred than in unbarred galaxies  
(Sakamoto \etal 1999). 
Once gas reaches the central regions of bars and 
builds up  large densities  comparable to the Toomre  critical density 
for the onset of gravitational instabilities, intense starbursts 
are triggered (Jogee \etal 2004a).
The frequency  of large-scale stellar  bars is in fact significantly 
higher in starburst galaxies than in normal galaxies (Hawarden \etal 96; 
Hunt \& Malkan 99),  while the question of whether Seyferts have an excess of 
large-scale bars remains under investigation  
(Regan \etal 1997;  Knapen \etal 2000;  Laurikainen \etal 2004). 
Although bars in the local Universe are fairly well studied, several 
fundamental aspects of  bar-driven  galaxy evolution remain 
open:

\begin{itemize}
%\vspace{-0.3 cm}
\item
When and how did bars form? 
Are they a recent (z$<$0.7) phenomenon or were they abundant 10 Gyr ago? 
Two conflicting results have been reported to date. 
Abraham \etal (1999)  claim a dramatic decline in the rest-frame 
optical bar fraction ($f_{\rm opt}$)  at $z>$~0.7, 
based on WFPC2 images of  a small sample of 
$\sim$ 50 moderately inclined spirals in 
the Hubble Deep Field (HDF). They find $f_{\rm opt}$  drops 
from $\sim$ 24\%  at $z\sim$~0.2--0.6  to below 5\%  at 
$z>$~0.7. A second study (Sheth \etal 2003), based on  NICMOS images
of the HDF, 
report an  \it observed \rm bar fraction of  5\% at  $z \sim$ 0.7--1.1.
Out of 95 galaxies in this redshift interval, 
they find four large bars with a mean semi-major axis $a$ of 
12 kpc  ($1.4\arcsec$). Since NICMOS has a large effective  PSF,
%   ($\ge$ 225 mas in a single image),  
this study fails to detect smaller bars  which,
at least locally, constitutes the majority of bars (see $\S$ 3). 
The authors  argue that their observed bar fraction of  5\%  
at $z >$ 0.7 is at least comparable to the local fraction  
of large (12 kpc)  bars, and thus, by extrapolation, the overall
optical bar fraction  over  all bar sizes (1.5-15 kpc)  probably
does not decline at $z >$ 0.7. 
This  conclusion suffers from  small number statistics 
and relies heavily on the extrapolation over bar sizes.

\item
%\vspace{-0.2 cm}
Are bars long-lived or do they dissolve and reform over a Hubble time?
Early studies  (e.g., Hasan \& Norman 1990; 
Norman, Sellwood, \& Hasan 1996) 
proposed that  once a  large  
central mass concentration (CMC) builds up in the inner 100 pc 
of a galaxy, for instance  via bar-driven gas inflow, it 
will destroy or weaken the bar by inducing 
chaotic orbits and reducing bar-supporting orbits.  
Subsequently, Athanassoula (2002) 
found  that in simulations with live halos, bars are more 
difficult to 
destroy due to resonant angular momentum exchange.
Shen \& Sellwood (2004)  find  that  in purely stellar
$N$-body simulations, the bar is  quite robust to CMCs.
%In some scenarios, bars  even cause their own demise and 
%self-destruction by  building up  such  CMCs via gas inflows. 
Furthermore, after a bar is destroyed by a CMC, the 
disk left behind is dynamically  hot, and does not readily reform 
a new bar unless it is cooled significantly.  
Recently, Bournaud \& Combes (2002) proposed that  
bars are destroyed primarily due to the reciprocal torques 
of gas on the stars in the bar  
(rather than by the CMC {\it per se}).  In their models, 
bars can dissolve and reform  recurrently 
over timescales $\le$ 3--4 Gyr, provided  the 
galaxy accretes  sufficient cold gas over  a Hubble time.

\item
%\vspace{-0.1cm}
Bar-driven evolutionary scenarios purport that bars may transform 
late-type spirals into intermediate types (e.g., Scd to Sbc) 
% byenhancing the  bulge-to-disk ($B/D$) ratios or  central mass
% concentrations in various ways. These include 
via  gas inflows which build 
%circumnuclear stellar disks and  
%high $V$/$\sigma$ 
pseudo-bulges,  and  via 
bending instabilities   (e.g., Sellwood 1993) or  vertical ILRs  
(e.g., Combes \etal 1990)   which drive stars to large scale
height into a   bulge-like configuration.
At intermediate redshifts, Katz \& Weinberg (2002) 
suggest bars can solve the dark matter halo cusp-core controversy.
While there is a wealth of circumstantial evidence 
(see review by Kormendy \& Kennicutt 2004) consistent 
with bar-driven  secular evolution, no systematic tests of this picture 
have been made either at local or intermediate redshifts. 
%of the impact of bars on  the structural evolution of galaxies.
\end {itemize}

\vspace{-0.1cm}
\noindent
Using $\sim$ 8,300 galaxies with superb $HST$ images and 
accurate redshifts over  $z \sim$~0.2--1.1  (corresponding to 
lookback times  $T_{\rm back}$ of $\sim$~2.5--8 Gyr) 
from the {\bf G}alaxy {\bf E}volution from {\bf M}orphologies and
{\bf S}EDs (GEMS;Rix \etal 2004; $\S$ 2)  survey, 
we have started  a comprehensive study of 
the properties (fractions, sizes, strengths)
of bars  as  a function of redshift, 
% out  to  $z\sim$ 1.1 
and host  galaxy properties.  
% (e.g., mass concentrations, sizes, activity),  
The GEMS bar project  will  constrain
how bars evolve  and 
impact the  structural evolution of galaxies over the last 8 Gyr, 
with the eventual goal of resolving some of 
the above open issues.
We describe here our methodology, early results, and upcoming  
highlights. 
These findings are described in more detail in Jogee 
\etal 2004b. 
We assume 
a flat cosmology with $\Omega_M = 1 - \Omega_{\Lambda} =0.3 $ 
and a Hubble constant $H_0$ = 70 km s$^{-1}$  Mpc$^{-1}$ 
throughout this paper. 

\vspace{-0.1cm}
\section[Imaging and Characterization Bars out to $z\sim$~1]
{Imaging and Characterization Bars out to $z\sim$~1 }

%\newpage
\begin{figure}[t]
\centerline{
\includegraphics[width=3.9in]{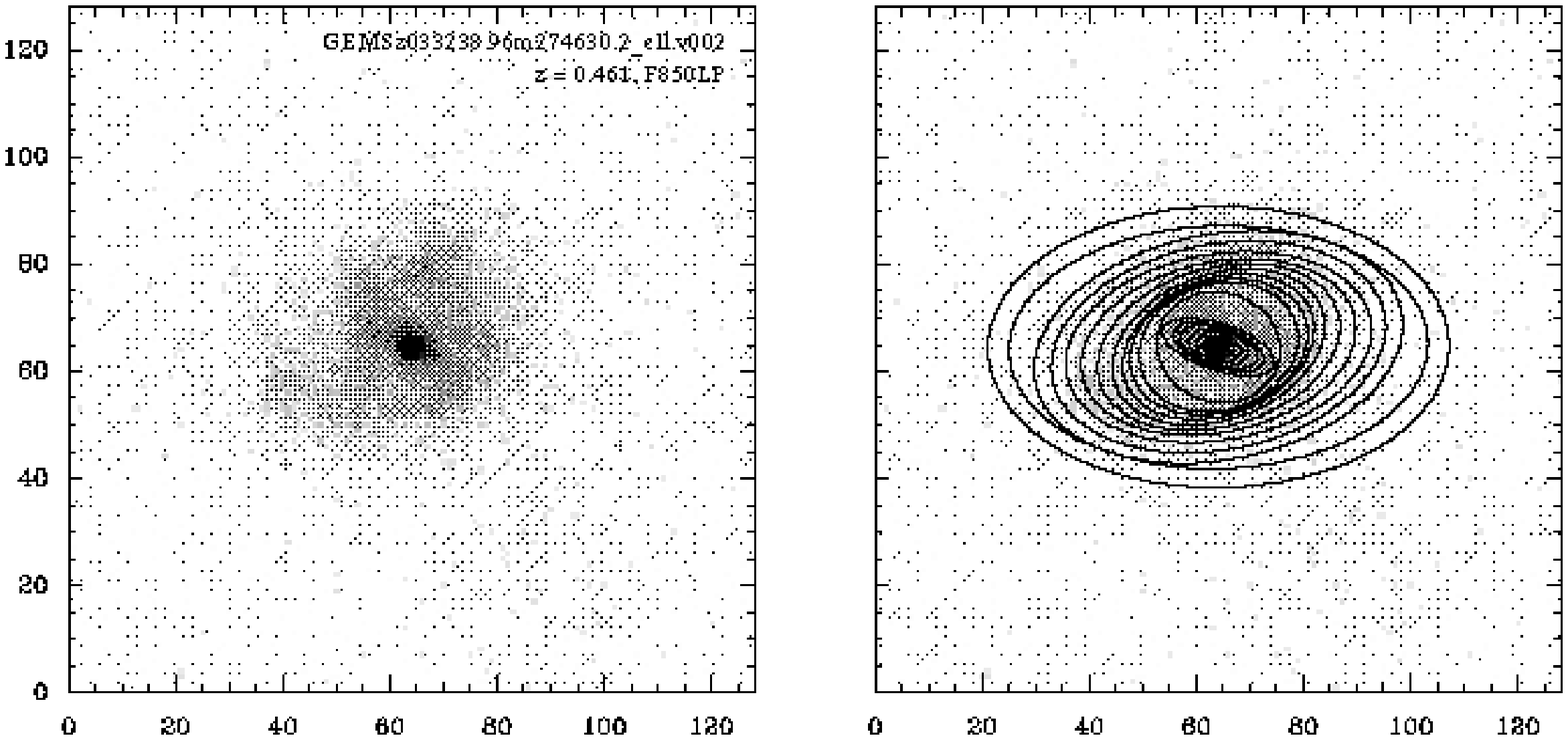}
}
\letteredcaption{a}{
\bf  Identification and characterization of bars at 
intermediate  redshifts --  \rm
The left panel  shows the  GEMS image of a  galaxy at $z \sim$~0.5 
(lookback time $T_{\rm back} \sim$ 5 Gyr) with a  bar, prominent 
spiral arms, and a disk. The right  
panel  show the same image   with an overlay of the
fitted isophotes.  The latter clearly  trace the bar, 
spirals, and outer disk. 
}
\end{figure}
\inxx{captions,lettered}

%\vspace{-3cm}
\begin{figure}[h]
\centerline{
\includegraphics[width=2.5in]{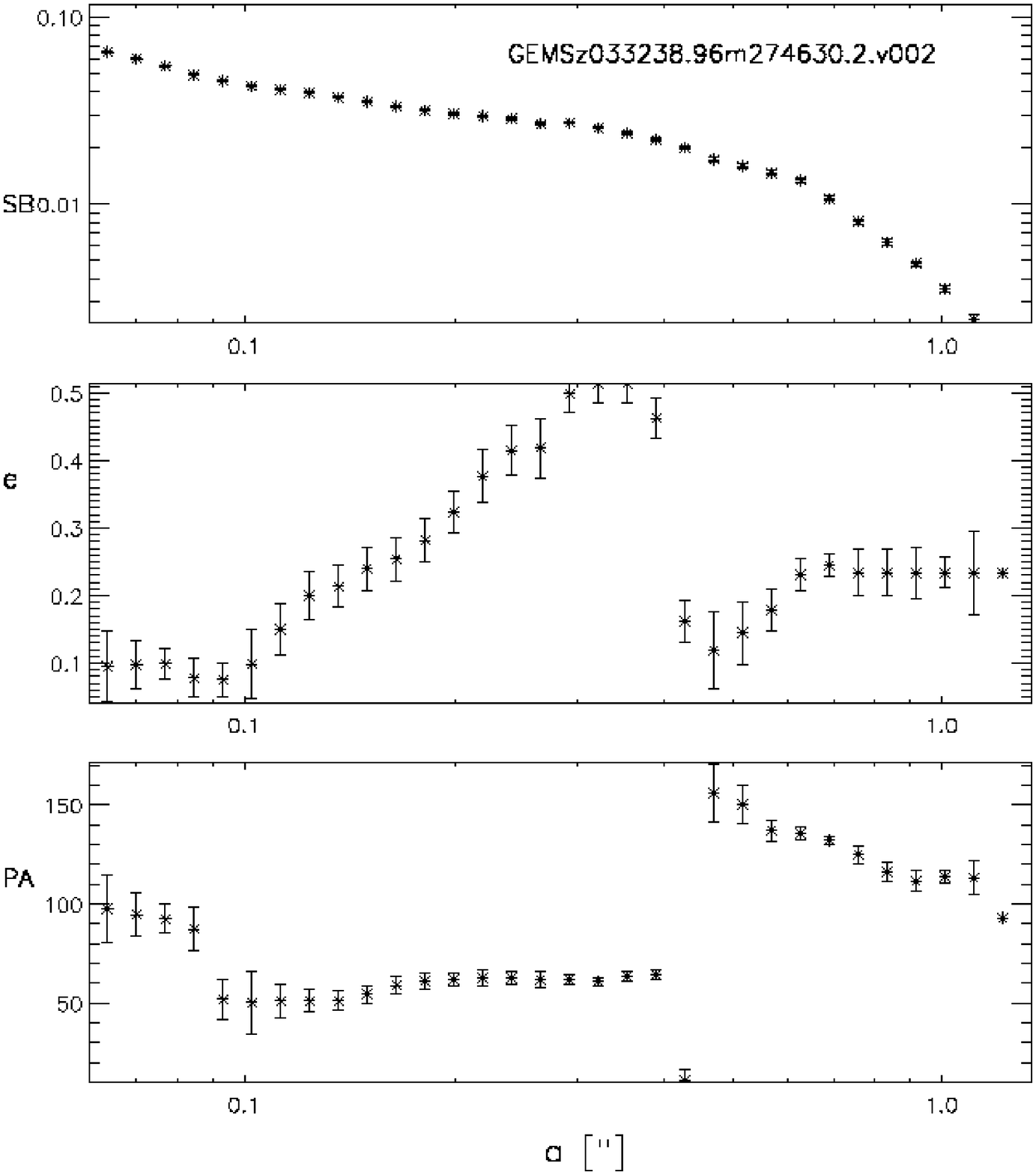}
}
\letteredcaption{b}{
\bf  Identification and characterization of bars at 
intermediate  redshifts --  \rm
The radial plots of the surface brightness (top),  
ellipticity $e$ (middle), and PA  (bottom)  generated 
by the isophotal fits in the previous figure are shown.
The  bar with a  semi-major axis $a\sim$~$0.31\arcsec$ 
causes $e$ to rise smoothly to a maximum, while the PA 
has a plateau over the region 
dominated by the $x_{\rm 1}$  orbits. At larger $a$, beyond the 
bar end,  $e$ drops sharply. The spiral arms 
lead to a twist in PA and varying $e$, and subsequently, the disk 
dominates. 
}
\end{figure}
\inxx{captions,lettered}

% The 
%remaining galaxies are classified as barred or unbarred and
%their  bar and disk parameters recorded.
%      Class =inclined   (i$>$60 deg) :  discard for sample 
%      Class= Barred , Unbarred
%      Class= D (need deprojection)

GEMS (Rix \etal 2004)  
is a two-color (F606W and F850LP) 
imaging survey with the $HST$  Advanced Camera for Surveys (ACS) 
of  a  large-area 
(800 arcmin$^2$  or $\sim 28\arcmin \times 28\arcmin$) field
centered on the $Chandra$ Deep Field South (CDF-S). 
The limiting 5$\sigma$ depth for compact sources is 
m$_{AB}$(F606W)=28.3 and m$_{AB}$(F850LP)=27.1. 
ACS imaging  from the GOODS project 
(Giavalisco et al. 2004) is used in the central quarter 
field. 
GEMS provides morphologies for  $\sim 8,300$  galaxies  
in the redshift range  $z \sim$ 0.2--1.1
where SEDs and  accurate redshifts 
($\sigma_{\rm z}$/(1 + $z$) $\sim$~0.02) down to  
$R_{\rm AB}$= 24   exist from 
the COMBO-17 project (Wolf \etal 2003). The dataset are further  
enhanced by panchromatic $Chandra$, $Spitzer$, and ground-based 
observations covering the X-rays to the far-IR and radio. 

The GEMS  ACS dataset offer many advantages for studying
bars, compared to earlier WFPC2/NIC3 imaging surveys.
The dataset provide a factor of 100 improvement 
in number statistics, an effective ACS PSF  which is 2--3 
times better than the  WFCP2 and NIC3 PSFs, and 
redder wavelength coverage and sensitivity  
via the ACS F850LP filter/CCD combination.
The superior PSF  allows us to probe bars 
down to small sizes ($0.12 \arcsec$ or 1 kpc; see $\S$ 3).
Finally, the large area ($\sim 120 \times$ HDF area) 
coverage reduces the effect of cosmic variance.

We identify and characterize bars out to $z\sim$~1.1 
 by performing isophotal  analyses of the 
F606W (\vb) and  F850LP (\zb)  images using an automated iterative
version of  of the IRAF STSDAS ``ellipse'' routine.   
% Our automated routine refines the center of an input image, 
% determines the maximum radius to fits based on the sky surface
% brightness 
Isophotal fits are  a good guide to the underlying orbital 
structure of a galaxy and provide  a  robust way of 
identifying and characterizing   bars  (e.g., Wozniak \etal 
1995; Jogee \etal 2002). 
A bar is identified by its  characteristic  signature in 
the radial profiles of ellipticity ($e$), surface brightness,  
and position angle  (PA), as illustrated in Figs. 1a and 1b.
The ellipticity is required to rise smoothly to a 
\it global \rm  maximum  above 0.25 while the PA 
has a plateau over the region dominated by  
the $x_{\rm 1}$  orbits,  followed by a drop $\ge $  0.1   in $e$
as the bar-to-disk transition occurs. 
The ellipticity, PA, and  semi-major axis for both the 
bar and outer disk can be identified from these profiles, 
thereby allowing the profiles to be 
subsequently deprojected to derive the intrinsic bar strength and
size. We also use the disk ellipticity and inferred inclination $i$
to reject  all  galaxies with $i <$~$60\deg$, since  
morphological  classification is unreliable in highly inclined galaxies.

\begin{figure}[ht]
\vskip.2in
\centerline{
\includegraphics[width=3.3in]{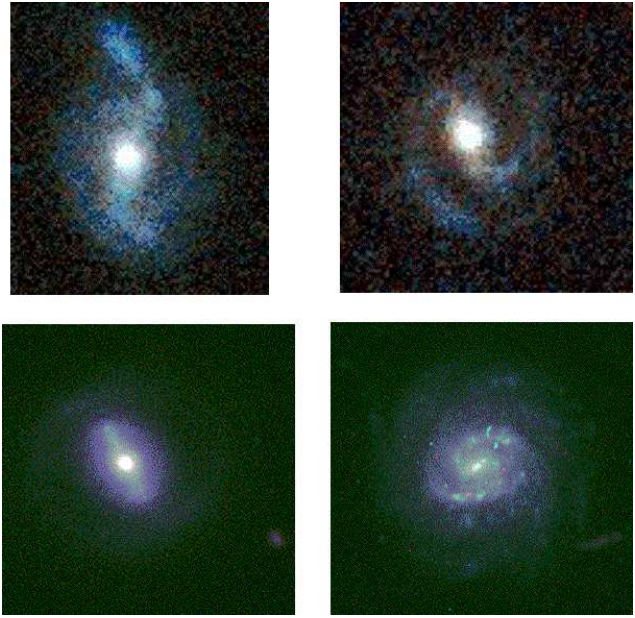}
}
\caption{Examples of bars and spiral arms in GEMS galaxies 
at $z\sim$ 0.3 and  0.4. (top panel:left to right), 
and  at $z\sim$ 0.5 and 0.9 (bottom panel:left to right) 
}
\end{figure}

\section[Early Results and Future Highlights]
{Early Results and Future Highlights}

We use two approaches to characterize the optical bar fraction 
$f_{\rm opt}$. 
The first is to identify bars in  the reddest observed band (F850LP) at 
all redshifts, with the idea that redder wavelengths are less impacted 
by extinction, and consequently better trace  the old stellar potential.
The disadvantage of this method  is that different, 
increasingly bluer rest-frame bands  are then traced in the 
two  redshift bins of interest  (0.24$<z \le$0.65 and 0.65$<z \le$1.1) 
as illustrated in Table 1a.  A complementary approach, illustrated
in Table 1b, is to use both F606W and F850LP images in order to 
work in a fixed rest-frame band ($B$/$V$)  and avoid
significant bandpass shifting.

We present early results based on a subsample of 600 galaxies 
distributed in the overlap area of GEMS and GOODS. 
Figure 2 illustrates  examples of  intermediate redshift 
galaxies with bars and spiral arms  in this sample. 
While being only a small fraction of the total 
GEMS dataset, this sample is 3--6 times larger than previous 
WFPC2 and NIC3 samples  used to study bars in the HDF.
An inclination cutoff of $i <$~$60\deg$  and a magnitude cutoff 
of  $R_{\rm AB} < 24$ mag are imposed in order to 
ensure accurate  photometric redshifts, spectral typing, and 
reliable morphological analyses.   Galaxies were assigned 
E/S0  and    (spiral+ starburst) types based on  comparing 
their  rest frame $U-V$  colors and absolute $M_{\rm V}$ magnitudes 
with  template SEDs  from Kinney et al (1996)  and 
Coleman, Wu, \& Weedman (1980).  We imposed an absolute 
magnitude cutoff of $M_{\rm V} < $ -19. The rest-frame optical 
bar fraction $f_{\rm opt}$  in a given redshift bin was  taken as 
($N_{\rm bar}$/$N_{\rm tot}$), where $N_{\rm bar}$ is the
number of bars detected, and $N_{\rm tot}$ is the  
number of spirals and starbursts. 
An alternative approach involves using the single fit Sersic index 
$n$ to identify disk galaxies as the main contributor to  $N_{\rm tot}$. 
Our exciting result, based on the first method, 
are highlighted below:

\begin{itemize}
\item
\it 
We find that the rest-frame optical bar fraction $f_{\rm opt}$  
remains constant at  $\sim$ ~30\%  over $z \sim$~0.2--1.1 
or lookback times  of $\sim$~2.5--9 Gyr. In particular, we do 
not find evidence of a dramatic decline in  $f_{\rm opt}$ to 
below 5\% at $z > 0.7$.
\rm
This result holds whether we  choose to use the reddest observed 
band (F850LP) in  both redshift bins (Table 1a),  or choose
to work in  the same  fixed rest-frame band (Table 1b).
To further test the robustness of the results, we repeated
the analyses  treating the  spiral  and  starburst types 
separately,  and found no change. 
We also performed an extra-consistency check by using the 
F606W   images over the  0.65$<z \le$1.1 interval 
to estimate the bar fraction ($f_{\rm uv}$) in the rest-frame UV. 
Strong bars are not expected to be delineated by recent 
massive SF as the strong shocks on their leading edges
are not  conducive to SF.  As expected, the 
estimated $f_{\rm uv}$ is much lower (5\%) than   $f_{\rm opt}$.
The results on $f_{\rm opt}$,  therefore, seem quite robust. 
Analyses of   a smaller sample  of ACS images of the
tadpole field is leading  to   similar conclusions (Elmegreen 
\etal 2004)

\begin{table}[ht]
\sidebyside 
{\letteredcaption{a}
{Optical bar fraction $f_{\rm opt}$ in the  reddest available filter (F850LP)}
\centering
\begin{tabular}{lcc}\sphline
% \it xxxxr  &  \it xxxx   &  \it xxxx \cr\sphline
Redshift &  0.24$<z \le$0.65  & 0.65$<z \le$1.1 \cr\sphline
Filter   &   F850LP           &  F850LP    \cr\sphline
Rest-frame   &  $I$/$V$	      &  $V$/$B$   \cr\sphline
Bar Fraction &   33\%        &    34\%      \cr\sphline 
\end{tabular}
\label{table1a}}
{\letteredcaption{b}
{Optical bar fraction $f_{\rm opt}$ in  a fixed rest-frame ($B/V$)}
\centering
\begin{tabular}{lcc}\sphline
% \it xxxxr  &  \it xxxx   &  \it xxxx \cr\sphline
Redshift &  0.24$<z \le$0.65  & 0.65$<z \le$1.1 \cr\sphline
Filter   &   F606W           &  F850LP    \cr\sphline
Rest-frame   &  $B$/$V$	      &  $V$/$B$   \cr\sphline
Bar Fraction &   25\%        &    34\%      \cr\sphline 
\end{tabular}
\label{table1b}}
\end{table}

\item
What kind of bars are we detecting at $z > 0.6$?
The bars at 
$z \sim$~0.6--1.3  have semi-major axes 
$a\sim$~$0.15\arcsec$--$1.8\arcsec$ and 1.6--14 kpc, and 
it is noteworthy that the 
majority of these bars have  $a$ below  $0.5\arcsec$  and 5 kpc,
and  would consequently be hard to detect without the superb resolution 
of ACS. 
The bars  have  rest-frame $B$-band 
ellipticities  $e_{\rm B}$ in the range 0.4--0.8 and  the
majority  have   $e_{\rm B} > 0.5$  (Fig.  3). 
\it 
The detected bars are thus  strong bars. 
\rm 
The fact that we not identify weak  (0.25 $< e_{\rm B} <$ 0.4) 
bars at $z \sim$~0.6--1.1  (Fig. 3) could lend itself to several  
interpretations.
One possibility  is that bars at $z \sim$ 1 were on average stronger 
than local ones (e.g., Bournaud \& Combes 2002).
Another  possibility is that cosmological dimming, loss of spatial 
resolution, and the potentially heavy  impact of dust and SF  
at  increasing redshifts  make it harder to 
detect such weak bars.  In order to quantify these effects 
we are artificially redshifting  local galaxies  out to  
$z \sim$~1.1  in rest-frame $B$-band, assuming  the current  
cosmology \it du jour, \rm  and folding in the ACS and GEMS
survey parameters. Weak short bars  are indeed hard to detect,  
but the exact effect of their non-detectability on the overall 
fraction is not yet fully quantified.

\begin{figure}[ht]
\vskip.2in
\centerline{
\includegraphics[width=2.1in]{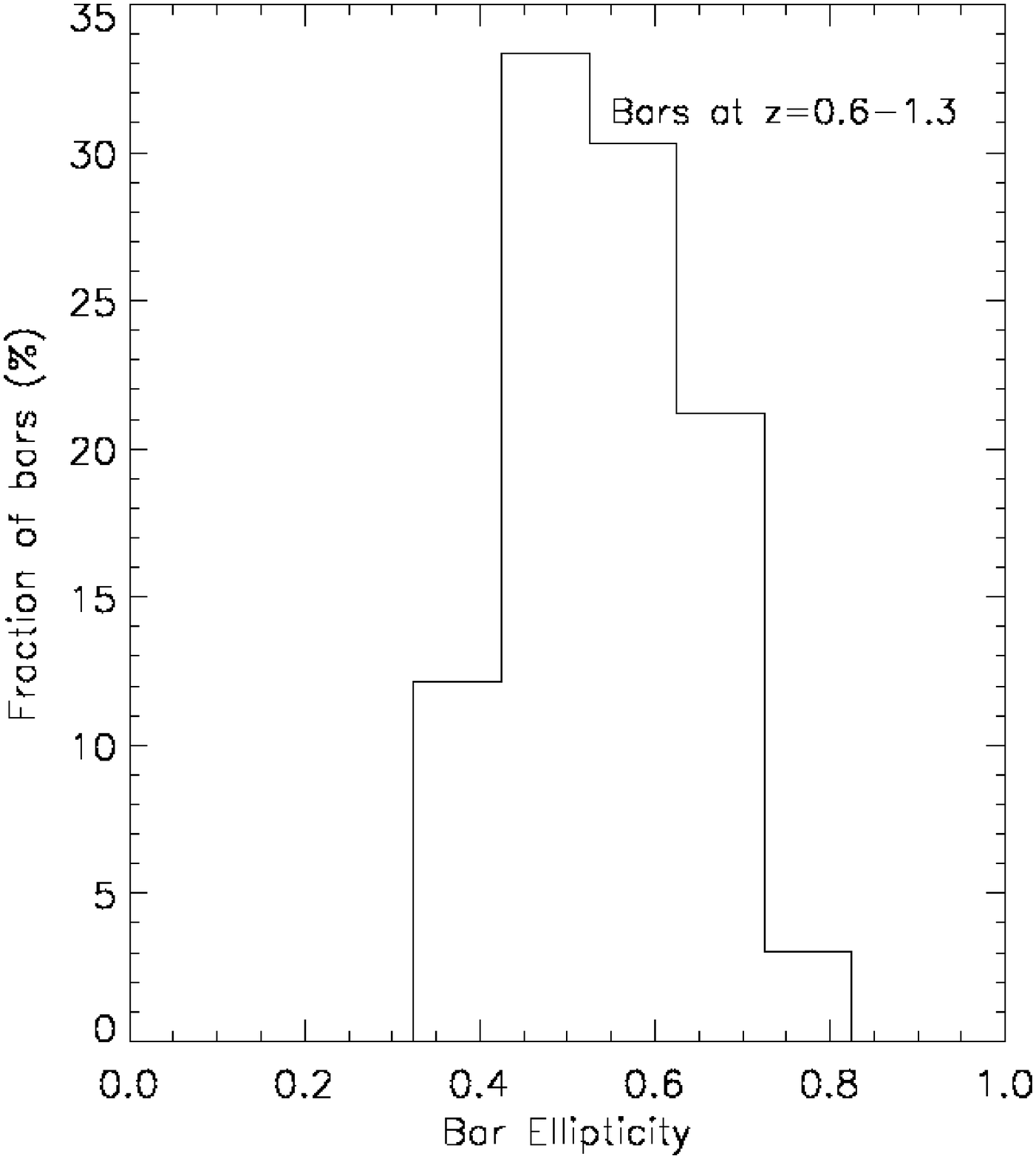}
}
\caption{
\bf Bars strengths  -- \rm 
The distribution of bar  strengths  (ellipticities  $e_{\rm B}$)  in the 
rest-frame $B$/$V$  is shown for bars  at $z \sim$~0.6--1.1,
over lookback times of 6--8  Gyr.
}
\end{figure}

\item
How do the bars  at $z > 0.6$ compare to local ones?
\rm 
\rm
Since carefully defined, large volume-limited samples of barred galaxies
for comparison  at $z\sim 0$  do not yet exist, we choose to compare the 
$z > 0.6$ bars to the local Ohio 
State University (OSU) sample of bright galaxies (Table 2; ES02)
\it 
It is remarkable to note 
that  strong ($e_{\rm B}$ =0.4--0.8)  bars which were  
in place 6--8 Gyr ago, and those at current epochs,  
have  a similar  rest-frame optical bar fraction  $f_{\rm opt}$  
\rm 
(Table 1b and 2)  
\it 
and a similar  bar size distribution 
\rm  
(Fig. 4). 
\rm 
The simplest naive interpretation of these results is that bars 
are long-lived over timescales $\ge$ 6 Gyr, and possibly over a 
Hubble time. However, we withhold any definite conclusions  
until we  have compared  
the normalized bar sizes (i.e., the ratio of bar size to disk 
size)  at these two epochs, and until we have increased our sample 
sufficiently so that the data can be binned over 1 Gyr interval 
over the last 9 Gyr.

\begin{table}[ht]
\caption
[Local bar fractions in the optical ($f_{\rm opt}$)  
and near-IR  ($f_{\rm NIR}$) based on the OSU sample
]
{Local bar fractions in the optical ($f_{\rm opt}$)  
and near-IR  ($f_{\rm NIR}$) based on the OSU sample
} 
\begin{center}
\begin{tabular*}{\textwidth}{@{\extracolsep{\fill}}lcccc}
%\begin{tabular}{ccccc} 
\hline
              &  Strong  &  Weak  & Strong+Weak  &  No  \cr
              &   Bars   &  Bars    &    Bars      & Bars \cr
\hline
% &  & & & \cr
% $f$ in $B$-band using RC3 bar strength$^a$ & 
%    35\%  \rm & 31\% & 66\%  & 34\% \cr
%and no  inclination cut-off     &  & & & \cr
$f$ in $B$-band using RC3 bar strength $^a$  & 
\bf  37\%  \rm & 34\% & 67\%  & 33\% \cr
%and inclination  $i < 60 \deg$      &  & & & \cr
$f$ in $B$-band using bar ellipticity  $e_{\rm B}$$^b$ & 
\bf 33\% \rm & 28\% & 61\% & 39\% \cr
%and inclination $i < 60 \deg$  &  & & & \cr
$f$ in $H$-band using bar ellipticity$e_{\rm H}$$^c$   &  
49\% & 21\% & 70\%  & 30\% \cr
%%and inclination $i < 60 \deg$    &  & & & \cr
\hline
\end{tabular*}
%\end{tabular}
\begin{tablenotes}
Notes to table --- The local bar fractions  in this table are based on the
OSU sample, after excluding all galaxies with inclination $i >$~60 $\deg$.
$a.$  The bar fractions shown here are from ES02  who use the  RC3 classes 
of SB, SAB, and SA  for ``strong bars'' , ``weak bars'',  and ``no bars''; 
% ES02 also find similar  fractions for the RC3 galaxy sample and  the 
% Carnegie Atlas of Galaxies;
$b.$  The bar fractions shown here 
assume  a $B$-band bar ellipticity $e_{\rm B} \ge $ 0.4  to define strong 
bars,  and   0.25~$\le e_{\rm B} <$~0.4 for weak bars;
$c.$  The bar fractions shown here  assume an $H$-band bar ellipticity 
$e_{\rm H} \ge $ 0.4  to define strong bars.
\end{tablenotes}
\end{center}
\end{table}
\inxx{captions,table}

\begin{figure}[ht]
%\vskip.2in
%\hspace{-0.3cm}
\centerline{%
\includegraphics[height=2.3in]{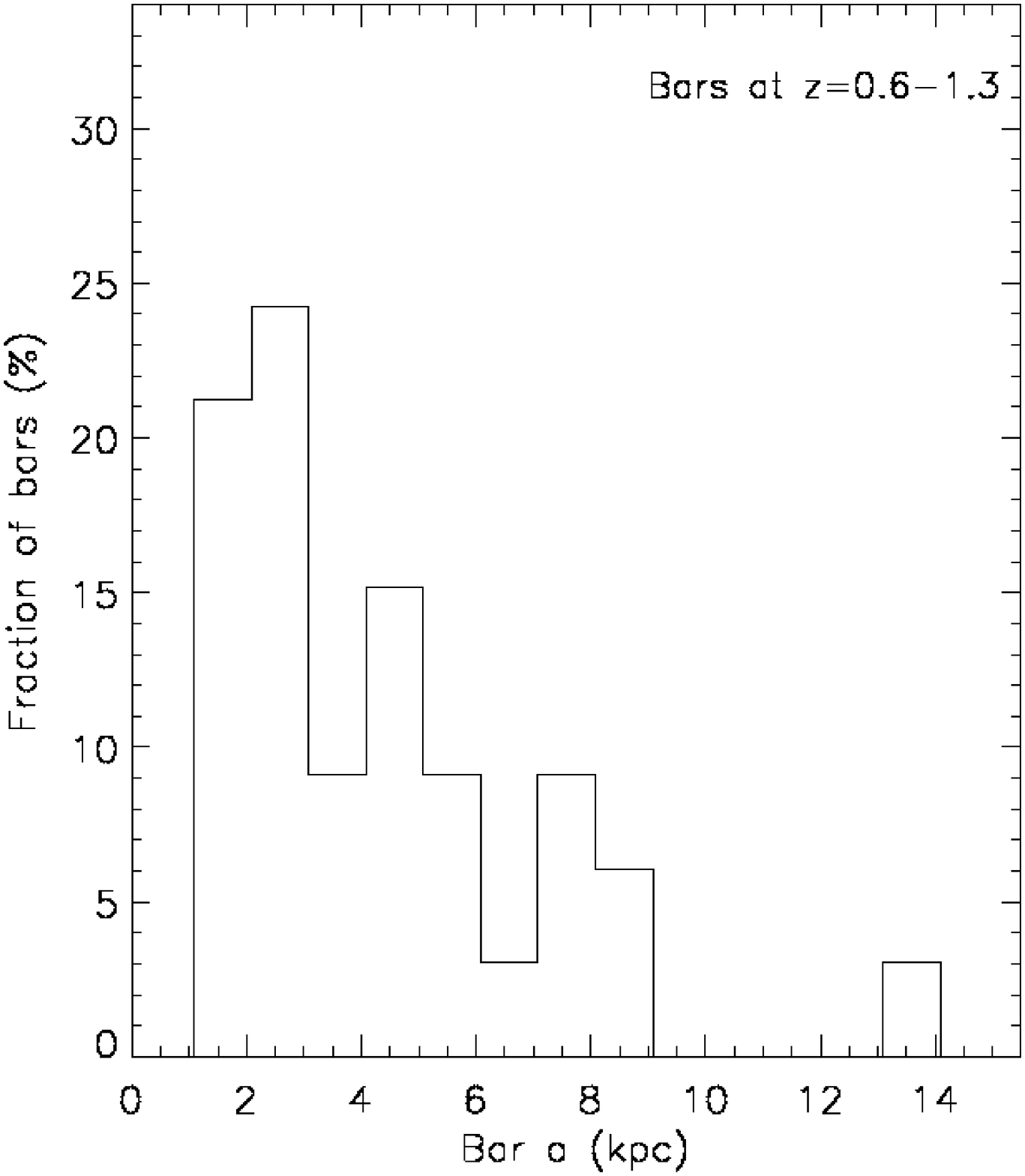}
\includegraphics[height=2.3in]{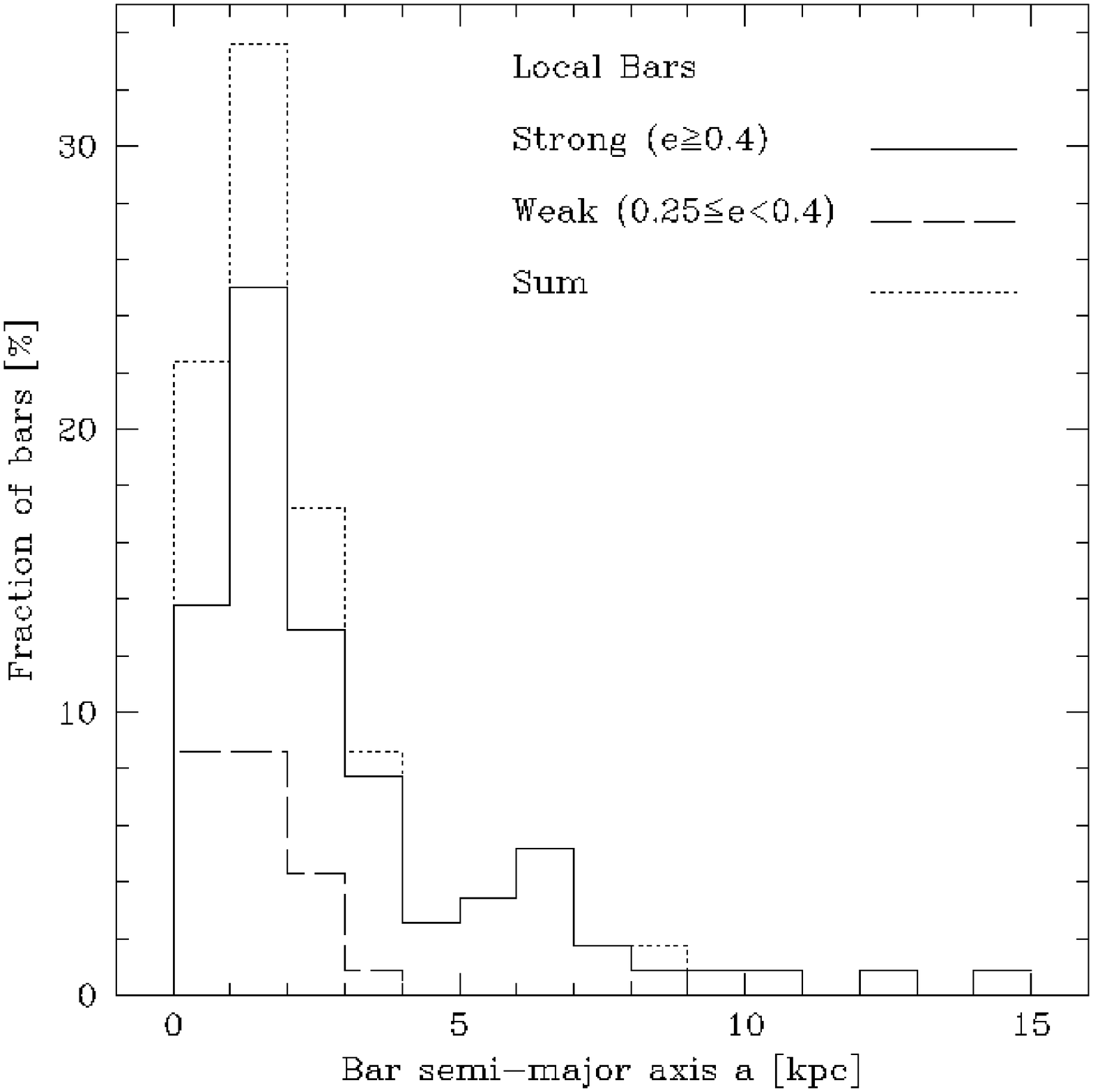}
}
\caption{
\bf  Distribution of bars sizes  6--8 Gyr ago compared 
to the present epoch -- \rm
$Left:$  \rm 
The distribution of bar semi-major axes $a$  in rest-frame $B$/$V$  
is shown for bars  at  $z \sim$~0.6--1.1  (lookback times 
$T_{\rm back}$ of $\sim$~6--8 Gyr) in the GEMS survey.
Note that most bars have $a<$ 5 kpc ($0.6\arcsec$ at $z \sim$~0.9)  
and would be difficult to detect without the exquisite 
resolution of ACS. 
$Right:$ \rm  Ditto,  but for local bars in the OSU sample after 
excluding inclined galaxies  with  $i> 60\deg $.  
The distribution is   shown for strong  
($e_{\rm B} \ge$ 0.4; solid line)  and weak
(0.25 $\ge e_{\rm B} <$ 0.4; hashed line) local bars. 
The strong local bars  are directly comparable to the bars 
detected at $z \sim$~0.6--1.1,  as the latter have  ellipticities 
$e_{\rm B} \ge$ 0.4.  There is a striking similarity 
in the size distribution of strong bars at the present epoch 
and 6--9 Gyr ago.
}
\end{figure}

\item 
\it 
Concluding remarks: 
\rm 
In recent years,  a fundamental puzzle was whether the  bar fraction shows a
dramatic decline at $z>$~0.6.   Mounting evidence (Jogee \etal 2004b, 
Elmegreen \etal 2004, Sheth \etal 2003) now suggests  this is not the case. 
The emphasis should now be to push investigations beyond mere  
estimates of  bar fractions.
The next challenge is  to  distinguish  between the  two main bar evolution 
scenarios: long-lived bars over a Hubble time versus  models of 
recurrent destruction/reformation of bars (e.g., Bournaud \& Combes 2002).  
In order to attack this  problem,  
we plan a systematic comparison of the properties of  bars (strength, 
normalized sizes) and host galaxies (central mass concentration, $B$/$D$, masses), 
over each  Gyr interval out to lookback times of 9 Gyr.
Another  important area to explore is the relationship of bars 
to starburst and AGN activity  out to $z \sim$~1, since the 
cosmic density of both types of activity show a dramatic increase 
from $z$=0 to 1.  The combination of GEMS ACS imaging, $Chandra$ 
observations,  and  inflowing  $Spitzer$ data makes these  timely issues 
to address, and hold the promise  of exciting science  ahead.

\end{itemize}

\begin{acknowledgments}
SJ   acknowledges support from  the National Aeronautics
and Space Administration (NASA) under  LTSA Grant  NAG5-13063
issued through the Office of Space Science, and thanks Paul 
Eskridge for kindly providing images and parameters for galaxies
in the OSU survey.
\end{acknowledgments}

\end{document}